\documentstyle[preprint,aps,prd,epsfig]{revtex}

\def\lprox{\mathrel{\raise .3ex\hbox{$<$\kern-
.75em\lower1ex\hbox{$\sim$}}}}

\def\gprox{\mathrel{\raise .3ex\hbox{$>$\kern-
.75em\lower1ex\hbox{$\sim$}}}}

\def\be{\begin{equation}}
\def\ee{\end{equation}}	
\def\ben{\begin{eqnarray}}
\def\een{\end{eqnarray}}

\begin{document}

\draft

\title{\bf Lightcone fluctuations in quantum gravity and extra dimensions}

\author{Hongwei Yu \footnote{e-mail: hwyu@cosmos2.phy.tufts.edu} 
        and L. H. Ford \footnote{e-mail: ford@cosmos2.phy.tufts.edu}}

\address{Institute of Cosmology, Department of Physics and Astronomy\\
	Tufts University, Medford, MA 02155, USA}

\date{\today}

\maketitle

\tightenlines

\begin{abstract}
We discuss how compactified extra dimensions may have potentially observable 
effects which grow as the compactification scale decreases. This arises because 
of lightcone fluctuations in the uncompactified dimensions which can result in 
the broadening of the spectral lines from distant sources. We analyze this 
effect in a five dimensional model, and argue that data from gamma ray burst
sources require the compactification length to be greater than about $10^5$ 
cm in this model. 

\end{abstract}



 

One of the most challenging problems in modern physics is the unification of the gravitational 
interaction with other known interactions in nature. Many attempts 
 involve going to higher dimensions and postulating the existence of extra 
spatial dimensions. 
If these  extra dimensions really exist, one must explain why they are not seen. The usual 
answer is that they curl into an extremely small compactified manifold, possibly
   as small as the Planck length,  $l_{pl}=1.6\times 10^{-33}$ cm. Therefore 
low-energy physics should be 
insensitive to them until distances of the compactification scale are being 
probed. In general, one has the possibility of observing the presence of the 
extra dimensions in a scattering experiment in which energies greater than that
associated with the compactification scale are achieved. Many extra dimension 
models are constained to have extra dimensions no larger than about
$ (1 {\rm Tev})^{-1} = 2 \times 10^{-17} {\rm cm}$.
 However, if only gravity
propagates in the extra dimensions, the upper bound can be much larger.  A 
recent proposal is that the fundamental scale of quantum 
gravity  can be as low as few Tev and the observed weakness of gravity is the 
result of large extra dimensions in which only gravity can propagate 
\cite{ADD,ADD2,ANT}.  
  The  size of extra dimensions (of which there must be at least 
two) can be as large as  1 mm in this type of model. 

 However, a question arises naturally as to whether there are any lower bounds 
on the sizes of extra dimensions. 
 It is the common belief that the existence of extra dimensions has no effect 
on low-energy physics
as long as  they are extremely small.
We will argue in this letter that this need not be the case.
The reason is lightcone fluctuations arising from the quantum
gravitational vacuum fluctuations, due to compactification of spatial 
dimensions\cite{YUF,Ford95}.  The compactification of spatial 
dimensions gives rise to stochastic fluctuations in the apparent speed of light 
which are in principle observable.  Basically,  the smaller the size of the 
compactified dimensions , the larger are the fluctuations that  result.
This is closely related to the Casimir effect,  the vacuum energy occuring 
whenever boundary conditions are imposed on a quantum
field.  The gravitational Casimir energy in the five-dimensional case with one 
compactified 
spatial dimension was studied in \cite{APCH}, where a nonzero energy density 
was found, which
tends to make the extra dimension  contract. This raises the question of 
stability of the extra
dimensions. It is possible, however, that the Casimir energy arising from the 
quantum 
gravitational field and other matter fields may be made to cancel each 
other\cite{RRT,Tsokos}, thus stabilizing the extra dimensions. Quantum lightcone 
fluctuations  due to the compactfication of
 spatial dimensions\cite{YUF}, although similar in nature to the Casimir effect,
 come solely from gravitons. Hence, no similar cancelation is to be 
 expected. 

Note that all of these quantum effects increase in magnitude as the
compactification scale decreases. The effect of compatification
in, for example, a five dimensional model can be thought of as producing
an infinite tower of massive modes with masses inversely proportional
to the compactification scale. One might think that the contribution of these 
modes to radiative correction would decrease with decreasing 
compactification scale, whereas in fact the reverse is
actually the case. Detailed calculations of the vacuum polarization \cite{F80}
and of the electron self-energy \cite{Y79} have been performed which reveal
growing effects with decreasing compactification length. This can be
understood as a consequence of the uncertainty principle; the fluctuations of
quantum fields confined in a finite region increase as the size of the
region decreases.

In a recent work, we studied the light cone fluctuations in  four 
dimensional flat spacetime
with compactification in one spatial dimension\cite{YUF}. It was found that
 these fluctuations, 
although typically of the order of the  Planck scale, can get larger for path 
lengths large compared
to the compactification scale. In particular, the mean deviation from the 
classical 
propagation time, $\Delta t$, is proportional to the square root of the travel 
distance, $r$.
 In this paper 
we apply the formalism to  spacetimes with extra dimensions, and
examine, in particular,  the five dimensional case 
 to demonstrate a possible observable 
consequence of  compactification of the extra dimension. 
  To begin, let us examine a $d$ dimensional spacetime with $d-4$  extra 
dimensions. 
 Consider a flat background  spacetime  with a linearized perturbation 
$h_{\mu\nu}$ propagating upon it , so the spacetime metric
may be written as  $
ds^2  = (\eta_{\mu\nu} +h_{\mu\nu})dx^\mu dx^\nu
= dt^2 -d{\bf x}^2 + h_{\mu\nu}dx^\mu dx^\nu \, ,  $
where the indices $\mu,\nu$ run through $0,1,2,3,...,d-1$.
Let $\sigma(x,x')$ be one half of the squared geodesic distance between
 a pair of spacetime points $x$ and $x'$,   and $\sigma_0(x,x')$ 
be the corresponding quantity in the flat background. 
In the presence of a linearized metric  perturbation
 $h_{\mu\nu}$, we may expand $
\sigma = \sigma_0 + \sigma_1 + O(h^2_{\mu\nu}) \, .$
Here $\sigma_1$ is first order  in $h_{\mu\nu}$. 
If we quantize  $h_{\mu\nu}$,
then  quantum gravitational vacuum
fluctuations will  lead to fluctuations in the geodesic separation, and 
therefore induce
 lightcone fluctuations.  In particular, we have 
$\langle \sigma_1^2 \rangle \not= 0$, since $\sigma_1$  becomes a quantum 
operator when the metric perturbations are quantized.   The quantum lightcone 
fluctuations give rise to 
stochastic fluctuations in the speed of light, which may produce an observable 
time delay or advance $\Delta t$ in the arrival times of pulses. Note that
this model uses a linearized approach to quantum gravity which is expected
to be a limit of a more exact theory. In the absence of a full theory,
this seems to be the most conservative way to compute quantum gravity effects.
One might comtemplate doing a one-loop calculation of an S-matrix element,
along the lines of those in Refs.~\cite{F80,Y79} for electrodynamics.
However, one would need to find a way to deal with the nonrenormalizability
of one-loop quantum gravity coupled to other fields \cite{DvN}. 

Let us  consider the propagation of light pulses between a source and a 
detector separated by a distance $r$ on a flat  background with 
quantized linear perturbations. In Ref.~\cite{Ford95} it was shown that
the root-mean-squared fluctuation in the propagation time is
\be
\Delta t = {\sqrt{\langle \sigma_1^2 \rangle_R}\over r} \, , \label{eq:delt1}
\ee
where $\langle \sigma_1^2 \rangle_R$ is a renormalized expectation value, which
was assumed to be positive. We can give an alternative derivation which applies 
in the case $\langle \sigma_1^2 \rangle_R < 0$.
For a pulse which is delayed or advanced by 
time $\Delta t$,
which is much less than $r$, one finds
\be
\sigma=\sigma_0+\sigma_1+....={1\over2}[(r+\Delta t)^2-r^2]\approx r\Delta t\,.
\ee
Take the fourth power of  the above equation and average over a given  quantum 
state of gravitons 
$|\phi\rangle$ (e.g. the vacuum states associated with  compactification of 
spatial dimensions), 
\be
\Delta t^4={\langle \phi| \sigma_1^4 |\phi\rangle\over r^4}\,.
\label{eq:TPHI}
\ee
This result is, however,  divergent due to the formal divergence of
 $\langle\phi| \sigma_1^4 |\phi\rangle $. 
 We can define $\langle \phi| \sigma_1^4 |\phi\rangle$ by normal ordering, 
and let 
 $
\langle \phi| \sigma_1^4 |\phi\rangle_R=
\langle \phi|:\sigma_1^4:|\phi\rangle\,$. 
For a free field $\psi$, Wick's theorem yields 
$ 
\langle:\psi^4:\rangle=3\langle:\psi^2:\rangle^2=3\langle \psi^2 \rangle_R ^2
$, where the expectation value is in the vacuum state.
Hence a suitable meaure of the  deviation from the classical propagation time is 
given by
\be
\Delta t = {(3\langle \sigma_1^4 \rangle_R)^{1/4}\over r}={3^{1/4}
\sqrt{|\langle \sigma_1^2 \rangle_R|}\over r}
\approx{\sqrt{|\langle \sigma_1^2 \rangle_R|}\over r}\,.
\label{eq:MDT}
\ee
Apart from small numerical factors which we ignore, this is equivalent to
replacing $\langle \sigma_1^2 \rangle_R$ by $|\langle \sigma_1^2 \rangle_R|$
in Eq.~(\ref{eq:delt1}).
The gauge invariance of this expression has been analyzed recently\cite{YUF}.
 
Note, however, that $\Delta t$ is the ensemble averaged deviation, not 
necessarily the expected variation in
flight time, $\delta t $,  of two pulses emitted close together in time. 
The latter is given by   $\Delta t$ only when two 
successive pulses are uncorrelated. 
This point is discussed in detail in Ref.~\cite{Ford96}.
These stochastic fluctuations in the apparent velocity of light arising from 
quantum gravitational 
fluctuations are in principle observable, since 
they may lead to a spread in the arrival times of pulses from
distant astrophysical sources, or the broadening of the spectral lines. 
This can be used to place a lower bound on the size
of the extra dimension. Lightcone fluctuations and their possible astrophysical 
observability have been recently discussed in a somewhat different framework 
in Refs.~\cite{NvD,AEMN,EMN}.

  In order to find $\Delta t$ in a particular situation, 
we need to calculate the quantum expectation value  $\langle \sigma_1^2 
\rangle_R$ in any chosen quantum state, which can be shown to 
be given by  \cite{YUF,Ford95}
\be
\langle \sigma_1^2 \rangle_R ={1\over 8}(\Delta r)^2
\int_{r_0}^{r_1} dr \int_{r_0}^{r_1} dr'
\:\,  n^{\mu} n^{\nu} n^{\rho} n^{\sigma}
\:\, G^{(1)R}_{\mu\nu\rho\sigma}(x,x') \,.
\label{eq:interval}
\ee
Here $ dr=|d{\bf x}|$,  $\Delta r=r_1-r_0$ and $ n^{\mu} =dx^{\mu}/dr$. The 
integration is taken along the null geodesic connecting two points $x$ and 
$x'$, and  
 $G^{(1)R}_{\mu\nu\rho\sigma}(x,x')$ is the graviton Hadamard  function,
 understood to be suitably renormalized. 
We shall now work  in the transverse-tracefree
 gauge in which the gravitational perturbations have only spatial components 
$h_{ij}$. 
This is a gauge which retains only physical degrees of freedom.  The quantized
field operator may be expanded as
\begin{equation}
h_{ij} = \sum_{{\bf k},\lambda}\, [a_{{\bf k}, \lambda} e_{ij} ({{\bf k}, 
\lambda})
  f_{\bf k} + H.c. ].
\end{equation}
Here H.c. denotes the Hermitian conjugate, $\lambda$ labels the ${1\over 2}
(d^2-3d)$ independent 
 polarization states,   $f_{\bf k}$  is the mode function, 
 and the $e_{\mu\nu} ({{\bf k}, \lambda})$
are polarization tensors. (Units in which $32\pi G_d =1$, where $G_d$ is
Newton's constant in d dimensions  and in which $\hbar =c =1$  will be used in 
this paper.)
Now suppose  the $(d-1)$-th dimension  is compactified into a 
small size of 
 periodicity length $L$,
so the mode function is given by
$
f_{\bf k} = (2\omega (2\pi)^{d-2}L)^{-{1\over 2}}  e^{i({\bf k \cdot x} -\omega 
t)}
	\label{eq:mode2}
$
with
$k_{d-1}={2\pi n\over L}, n=0,\pm 1, \pm 2, \pm 3,...\,.$ 
Let us denote the associated vacuum state by $|0_L\rangle$. 
 In order to calculate the 
gravitational vacuum fluctuations due to compactification of the extra
 dimension, we need the 
renormalized graviton Hadamard function with respect to the vacuum state 
$|0_L\rangle$, 
$G^{(1)R}_{ijkl}(x,x')$, which can be seen to be given by an image sum of the 
corresponding Hadamard function for the uncompactified Minkowski  
 vacuum, $G_{ijkl}^{(1)}$ :
\be
 G_{ijkl}^{(1) R}(t, x_{d-1},\,t',x_{d-1}')
 ={\sum_{n=-\infty}^{+\infty}}^{\prime}G_{ijkl}^{(1) }(t,x_{d-1}, \,t',x_{d-1}' 
+nL)\, ,
\ee
where the prime on the summation indicates that the $n=0$ term is excluded and 
the notation 
$(t, x_1,.., x_{d-2}, x_{d-1})\equiv(t, x_{d-1}) $ has been adopted. 

  From now on, we will restrict ourselves to the five dimensional case with one 
compactified dimension. Because the model discussed in this paper has only 
one extra dimension, the results do not bear directly on the large extra 
dimension models \cite{ADD,ADD2}, which require at least two extra dimensions.
Higher dimensional cases will be discussed in detail 
elsewhere~\cite{YUF2}. 
To see how light cone fluctuations arise in the usual uncompactified space 
as a result of compactification of the extra dimension,
let us consider a light ray traveling along the
$x_1$ direction from point $a$ to point $b$, which is perpendicular to the 
direction of 
compactification. The relevant graviton two-point function is $G_{xxxx}$, 
which can 
be shown to be given by 
\ben
G_{xxxx}(t,x_4,\, t',x_4')&=&2[ D(t,x_4,\, t',x_4')-2F_{xx}(t,x_4,\, t',x_4')
\nonumber\\
&&+H_{xxxx}(t,x_4,\, t',x_4')]\,.
\een
Here $D(x,x')$,  $F_{ij}(x,x')$ and $H_{ijkl}(x,x')$ are functions given by.
\ben
 	D(x,x')&=&
{Re\over{(2\pi)^4}}\int\, {d^4{\bf k}\over{2 \omega}}e^{i{\bf k} 
\cdot({\bf x}-{\bf x'})}e^{-i\omega(t-t')}\nonumber\\
&&={1\over 8\pi^2}{1\over (R^2-\Delta t^2)^{(3/2)}} \,,
\een

\ben
 	F_{ij}(x,x')&&={Re\over{(2\pi)^4}}\partial_i\partial'_j\int\, 
{d^4{\bf k}\over{2 \omega^3}}e^{i{\bf k} \cdot({\bf x}-{\bf x'})}
e^{-i\omega(t-t')}\nonumber\\
&&={1\over{8\pi^2}}\partial_i\partial'_j\, 
\left({\sqrt{R^2-\Delta t^2}\over R^2}\right)
 \,,
\een
and 
\ben
 	H_{ijkl}(x,x')&&={Re\over{(2\pi)^4}}
   \partial_i\partial^{\prime}_j\partial_k\partial_l^{\prime}
\int\, {d^4{\bf k}\over{2 \omega^5}}e^{i{\bf k} \cdot({\bf x}-
{\bf x'})}e^{-i\omega(t-t')}
\nonumber\\
&&=0\,.
\een
Here $R=|{\bf x-x'}|$ and $\Delta t=t-t'$.

 Inserting the above results into Eq.~(\ref{eq:interval}),
carrying out the differentiation in the function $F_{x_1x_1}$, using the fact 
that $x_1-x_1'=\Delta t$,  and then performing the integration, we finally find
\be
\langle \sigma_1^2 \rangle_R ={r^2\over 32\pi^2L}{\sum_{n=1}^{\infty}}\,
\left [
{8\over n}\ln\bigl(1+{\rho^2\over n^2}\bigr)-{2\rho^2\over n^3}-{8\rho^2\over
(\rho^2+n^2)n}\right]\,.
\ee
Here we have defined $r=a-b$ and a dimensionless parameter $\rho=r/L$.  We are 
interested  in the case in which
$\rho \gg 1$. It then follows that the summation is dominated, to the leading 
order, by
the second term, 
\be
\langle \sigma_1^2 \rangle_R \approx - 
{r^2\over 16\pi^2L} {\sum_{n=1}^{\infty}}\,{\rho^2\over n^3}
 = -{\zeta(3) r^2\rho^2\over 16\pi^2L}\,,
\ee
where $\zeta(3)$ is the Riemann-zeta function.
So, the mean deviation from the classical propagation time due to the lightcone 
fluctuations
is
\be
\Delta t \approx \sqrt{ {\zeta(3)\over 16\pi^2L}}\sqrt{32\pi G_5}\,\rho
\approx \biggl({ r\over L}\biggr)\,t_{pl}\,.
\label{eq:t1}
\ee
Here we have used the fact that $G_5=G_4\, L$, and   $t_{pl} \approx 5.39\times 
10^{-44}
s$ is the Planck time.

Here $\Delta t$ increases linearly with the propagation
distance, in contrast to the square root growth found in four dimensional 
compactified spacetime \cite{YUF}.
Equation~(\ref{eq:t1}) also reveals that the lightcone fluctuation effect is 
inversely related to $L$, the compactification length. This is due to the 
increased quantum fluctuations of fields confined in a smaller region. When
$r$ is of cosmological dimensions and $L$ is sufficiently small, the effect
is potentially observable.

Before we proceed further, it should be noted that we have set 
$\langle \sigma_1^2 \rangle_R =0$ when $L \rightarrow \infty$. This is the
most natural choice of renormalization, corresponding to the effect of the
graviton fluctuations vanishing in the limit of noncompactified spacetime.
This is analogous to setting a Casimir energy density to zero in the limit
of infinite plate separation.
There is, however, another logical possibility.
This is to set $\langle \sigma_1^2 \rangle_R =0$ at a finite value of $L$.
If $L$ is constant, then this procedure removes the lightcone fluctuations
and sets $\Delta t =0$. In our view, this is an unsatisfactory solution,
as there seems to be nothing in the theory which picks out a particular
finite value of $L$. In any case, changes in $L$ would still be observable,
as one could at most set $\langle \sigma_1^2 \rangle_R =0$ at one point
along the path of a light ray. Let $L_i$ be the compactification length at
the time of emission, and $L_f = L_i(1 +\delta)$, where $|\delta| \ll 1$,
be that at the time of detection. Then the variation in flight times must 
satisfy~\cite{YUF2}
\be
\Delta t \agt \sqrt{|\delta|}\, \biggl({ r\over L}\biggr)\,t_{pl} \,.
\label{eq:Tvari}
\ee
The limits on the time variation of fundamental constants place various upper
bounds \cite{KPW,JDB} on $|\delta|$ in the range between $10^{-2}$ 
and $10^{-10}$.

The variation in the flight time of pulses, $\Delta t$, can apply to the 
successive wave crests of a plane wave. This leads to a broadening of spectral 
lines from a distant source. Note, however, that  $\Delta t$ is the
 expected variation in the arrival times of two successive pulses only when 
they are uncorrelated.
Following
Ref.~\cite{Ford96}, we have examined~\cite{YUF2} the correlation between 
two successive pulses separated in time by $T$, and found that if $r \gg T, L$,
the   
two pulses are uncorrelated when $T\gg L^2/r$. Thus the pulses can be 
uncorrelated 
even when $T \ll L$. Now
suppose that the
 experimental fractional resolution for a particular spectral line of period 
$T$ is $\Gamma$. Then we must have ${\Delta t/ T}\leq \Gamma$, which leads 
to a lower bound on $L$ of
\be
L\agt {r\,t_{pl}\over \Gamma T} = \frac{4\times 10^4 cm}{\Gamma} \,
\left(\frac{r}{1000 Mpc}\right)\, \left(\frac{E}{1 MeV}\right) \,,
\label{eq:Lbound}
\ee
where $E$ is the energy of a photon with period $T$.
This bound can be trusted only when the condition for uncorrelated pulses,
\be
L\ll \sqrt{rT} = 6\times 10^8 cm\, \left(\frac{r}{1000 Mpc}\right)^\frac{1}{2}
\, \left(\frac{1 MeV}{E}\right)^\frac{1}{2}
\,,  \label{eq:uncorr}
\ee
is satisfied.

The strongest lower  bound would  be deduced from the highest frequency 
spectral  lines of the most distant sources, with the highest observed 
resolution. The best compromise between these various requirements seems to be
data from gamma ray burst sources, which involve cosmological distances and
high frequencies, albeit low resolution. The use of gamma ray burst to 
constrain quantum gravity models was discussed by Amelino-Camelia, {\it et al}
\cite{AC}. A typical burst is GRB990123, which involved\cite{GRB990123}
gamma rays at an energy of the order of $1\, MeV$ from a source
with a redshift of at least $z = 1.6$. Assuming a Hubble constant of 
$H_0 = 65 km/s/Mpc$ and a matter dominated universe, this corresponds to a 
distance of $r \agt 2400\, Mpc$. If we take the resoluton $\Gamma$ to be of 
order unity, this leads to
a lower bound on $L$ from Eq.~(\ref{eq:t1}) of $L \geq 10^5 cm$. This lower 
bound satisfies Eq.~(\ref{eq:uncorr}). This is a remarkably strong bound
which would seem to rule out the five dimensional theory. Even if one adopts
the approach of only using Eq.~(\ref{eq:Tvari}) to bound $|\delta|$, the 
result is $|\delta| \leq 10^{-10} (L/ 1\, cm)^2$, which for $L \ll 1 cm$
is much stronger than
the bounds cited above based upon time dependence of fundamental constants.
It is of interest to note that reasonably strong bounds on $L$ can be 
obtained from much lower frequency sources. Microwave lines from quasars
\cite{YUF2}, and the cosmic microwave background \cite{DFY} both yield lower 
bounds on $L$ of the order of a few tenths of a millimeter.

To conclude, we have demonstrated, in the case of one extra dimension, that
the large quantum lightcone fluctuations due to the compactification of the 
extra dimension require the size of the extra dimension to be macroscopically 
large. This result seems to rule out the five dimensional Kaluza-Klein theory,
or at the very least, place strong limits on the rate of change of the extra
dimension. We must point 
out that the rate of growth of $\Delta t$ with $r$  depends crucially on the 
number and nature of the 
spatial dimensions. In four dimensions, $\Delta t\propto \sqrt{r/L}$, 
while in five dimensions 
$\Delta t \propto r/L$. We have also analyzed six and higher dimensional models
in which the extra dimensions are flat. Here we find \cite{YUF2} 
that $\Delta t$ grows
only as a  logarithmic function of $r/L$. In these models, data on spectral
lines from distant sources yield no significant constraints on the 
compactification scale. 

In principle, the effect discussed in this paper should be a generic
phenomenon to be expected whenever there are small extra dimensions. This 
follows from the uncertainty principle argument given above, to the effect
that confining the graviton modes in a small region of space should give rise
to large fluctuations inversely related to the size of the region. However,
there is a possibility of subtle cancellations making the net effect smaller
than would naively be expected. This is what happens in the models with two or
more flat extra dimensions.
However, with two or more extra dimensions, it would 
also be possible to have the extra dimensions curved. The lightcone fluctuations
in such models have not yet been examined. It is entirely possible that they 
might exhibit a rate of growth intermediate between linear and logarithmic
functions. The phenomenon discussed in this 
paper can not only constrain models with extra dimensions, but could 
conceivably lead to positive confirmation of the existence of such dimensions.

Acknowledgment:
We would like to thank R. Di Stefano and W. Waller for useful discussions. 
This work was 
supported in part by the National Science Foundation under Grant PHY-9800965.


\begin{references}

\bibitem {ADD} N. Arkani-Hamed, S. Dimopoulos and G. Dvali,  Phys. Lett. 
{\bf B429} (1998) 263.

\bibitem {ADD2} I. Antoniadis, N. Arkani-Hamed, S. Dimopoulos and G. Dvali, Phys. Lett. 
{\bf B436} (1998) 257;  N. Arkani-Hamed, S. Dimopoulos and G. Dvali, Phys. Rev. D
{\bf 59} (1999) 086004.

\bibitem {ANT} I. Antoniadis, Phys. Lett. {\bf B246} (1990) 377.

\bibitem{YUF} H. Yu and L.H. Ford, Phys. Rev. {\bf D60} (1999) 084023,
 gr-qc/9904082.

\bibitem{Ford95} L.H. Ford, Phys. Rev.  {\bf D51} 
               (1995) 1692, gr-qc/9410043. 


\bibitem{APCH} T. Appelquist and A. Chodos,  Phys. Rev. 
{\bf D28}  (1983) 772.

\bibitem{RRT} M.A. Rubin and B. Roth, Phys. Lett. {\bf 127B} (1983) 55.
 
\bibitem{Tsokos} K. Tsokos, Phys. Lett. {\bf 126B} (1983) 451. 

\bibitem{F80} L.H. Ford, Phys. Rev.  {\bf D21} (1980) 933.

\bibitem{Y79} T. Yoshimura,  Phys. Lett. {\bf 72A} (1979) 391.

\bibitem{DvN} S. Deser and P. van Nieuwenhuizen, Phys. Rev. {\bf D10}
(1974) 401, 411.

\bibitem{Ford96} L.H. Ford and N. F. Svaiter, Phys. Rev.  {\bf D54} 
               (1996) 2640, gr-qc/9604052.

\bibitem{NvD} Y. J. Ng and  H. van Dam, Mod. Phys. Lett. A {\bf 9} (1994)
335.  

\bibitem{AEMN} G. Amelino-Camelia, J. Ellis,  N.E. Mavromatos and 
D.V. Nanopoulos, Int. J. Mod. 
Phys.  {\bf A12}  (1997) 607. 


\bibitem{EMN} J. Ellis, N.E. Mavromatos and D.V. Nanopoulos, 
Phys.Rev. D {\bf 61} (2000) 027503, gr-qc/9906029. 

\bibitem{YUF2} Hongwei Yu and L.H. Ford, gr-qc/0004063. 

\bibitem{KPW} E.W. Kolb, M.J. Perry, and T.P. Walker, Phys. Rev.  {\bf D33}
 (1986) 869. 

\bibitem{JDB} J.D. Barrow, Phys. Rev.  {\bf D35} (1987) 1805. 

\bibitem{AC} G. Amelino-Camelia, J. Ellis,  N.E. Mavromatos, 
D.V. Nanopoulos, and S. Sarkar, Nature {\bf 393} (1998) 763.

\bibitem{GRB990123} S.R. Kulkarni, {\it et al}. Nature {\bf 398} (1999) 389.

\bibitem{DFY} R. DiStefano, L.H. Ford, and H. Yu, manuscript submitted to
Astrophysical J.

\end{references}
\end{document}